\documentclass[fleqn,usenatbib,useAMS]{mnras}
\usepackage[T1]{fontenc}
\usepackage{ae,aecompl}
\usepackage{newtxtext,newtxmath}
\usepackage{graphicx}
\usepackage{amsfonts}
\usepackage{amssymb}
\usepackage{colors}
\usepackage[normalem]{ulem}

\title[Dark galaxies]{The distribution of dark galaxies and spin bias}
\author[R. Jimenez, A. F. Heavens]{Raul Jimenez$^{1,2}$\thanks{raul.jimenez@icc.ub.edu}, Alan F. Heavens$^3$\thanks{a.heavens@imperial.ac.uk}\\
$^1$ICC, University of Barcelona, Marti i Franques 1, 08028, Barcelona, Spain.\\
$^2$ICREA, Pg. Lluis Companys 23, 08010 Barcelona, Spain.\\
$^3$Imperial Centre for Inference and Cosmology (ICIC), Department of Physics,  Imperial College London, London SW7 2AZ\\
}

\newcommand{\be}{\begin{equation}}  \newcommand{\ee}{\end{equation}}
  \newcommand{\ba}{\begin{eqnarray}}
\newcommand{\ea}{\end{eqnarray}}

\def\gs{\mathrel{\raise1.16pt\hbox{$>$}\kern-7.0pt %
\lower3.06pt\hbox{{$\scriptstyle \sim$}}}}         %
\def\ls{\mathrel{\raise1.16pt\hbox{$<$}\kern-7.0pt %
\lower3.06pt\hbox{{$\scriptstyle \sim$}}}}         %

\begin{document}

\voffset=-0.25in 

\maketitle

\begin{abstract}
In the light of the discovery of numerous (almost) dark galaxies from the ALFALAFA and LITTLE THINGS surveys, we revisit the predictions of the existence of dark galaxies, based on the Toomre stability of rapidly-spinning gas disks.  We have updated the predictions for $\Lambda$CDM with parameters given by the Planck18 collaboration, computing the expected number densities of dark objects, and their spin parameter and mass distributions.  Comparing with the data is more challenging, but where the spins are more reliably determined, the spins are close to the threshold for disks to be stable according to the Toomre criterion, where the expected number density is highest, and reinforces the concept that there is a bias in the formation of luminous galaxies based on the spin of their parent halo.
\end{abstract}


\section{Introduction}
\label{Introduction}

Star formation from the interstellar medium requires a dramatic collapse of gas, from a density $\sim 10^{-20}$~kg m$^{-3}$ to $\sim 10^3$~kg m$^{-3}$ in the case of the Sun.  Such a large increase in density may be a very difficult task if the surface or volume density of gas in a disc of a galaxy is low enough, and it is not implausible that in certain situations it may be impossible to reach the density needed to form stars.  Detailed physical arguments support this notion, going back to \cite{Toomre64}, who found that a natural threshold should to exist for triggering instabilities in a rotating disk.  This is essentially the Jeans stability but now including the fact that  the rotating disk shear adds effectively to the outward pressure. In its simplest form, for an infinitesimal thin disk, the onset of instability is determined by the dimensionless quantity
$Q = (\kappa c_s)/(\pi G \Sigma) \lesssim 1$, where $\Sigma$ is the surface density, $G$ Newton's constant, $c_s$ the sound speed and $\kappa$ the epicyclic frequency. More general formulations of the Toomre criterion for thick disks or infalling material, amount to corrections order one of the original Toomre's formula~\citep{Rafikov}.   

In the intervening years, support for the Toomre criterion has been found in thin disks \citep{MartinKennicutt01,MKH02,Genzel10,RomeoMogotsi17}, but \cite{Elson12} suggest that it may not apply in dwarf irregulars, and there are suggestions that magnetic fields \citep{Takahira18} or cloud-cloud collisions \citep{Aouad03} could also play a role.   However, there are a number of theories that aim to explain why $Q\simeq 1$ for star-forming disks (see the discussion in \cite{Krumholz18} and references therein).  Although there is a debate about whether the Toomre criterion is the full story when the disk is unstable, where mechanisms to keep the disk marginally unstable seem required (e.g.  \cite{KrumholzBurkert10,Krumholz18}), our focus here is on the opposite case, where no stars are forming, and the criterion $Q\gs 1$ is more secure and we assume that it applies.  In this letter, we investigate the properties of the HI-rich gas disks discovered by \cite{Leisman} from the ALFALFA survey and by \cite{Butler} from the LITTLE THINGS survey, and look to see if they are consistent with updated predictions of gas stability, based on the Toomre $Q$ value and the $\Lambda$CDM model.

One route to decrease the surface density of the disk is by  higher values of the initial spin of the dark halo where it settles. In the hierarchical $\Lambda$CDM model of structure formation \citep{Planck18}, where cold dark matter dominates the potential wells, galaxies obtain their torques because of tidal forces from the density field~\citep{Doro,HP88}. For an initially  gaussian field of primordial fluctuations, it is possible to predict the distribution of the halo spin parameter $\lambda = J |E|^{1/2}/GM^{5/2}$, where $J$ is the angular momentum of the halo and $E$  and $M$ its energy and mass.   The  result of the calculation in \cite{darkgal} was that for masses below a given threshold all disk galaxies should be almost dark, i.e. the disks were Toomre stable throughout.

By the time of the prediction by \citet{darkgal} there were no dark galaxies observed. Indeed, we pointed out that "A firm prediction of our model is that [dark galaxies] should be seen in deep HI surveys in voids." After more than 20 years, there has been a revolution in the coverage and sensitivity of blind HI surveys. In the same vein that we have witnessed a revolution in optical surveys, a similar trend has occurred for HI surveys.  Recent HI surveys have indeed revealed a number of dark (or near dark) galaxies. The purpose of this short note is to update the predictions to $\Lambda$CDM, including number densities, and deriving the joint spin-mass distribution, and to analyse if the newly found HI-rich galaxies parameters are consistent with the model. Our main conclusion is that HI (almost) dark galaxies properties do indeed resemble the ones predicted, and their number density is consistent with them being the tail of the spin distribution in hierarchical tidal-torque theory as in the $\Lambda$CDM model.

\section{Theory}
\label{Theory}

As in \citet{darkgal}, we model galaxies as disks. Assuming that the threshold for star formation is described by a Toomre stability parameter $Q=\Sigma_c/\Sigma$, where $\Sigma_c = \alpha \kappa c_{\rm disp}/(3.36 G)$, with $\alpha \sim 1$, $c_{\rm disp}$ is the velocity dispersion of the disc and $\kappa = 1.41 (v/r) \sqrt{1+d \ln v/d \ln r}$, $v$ the rotational velocity of the disc and $r$ the distance from the centre of the disc, it is possible to compute for which masses and values of $\lambda$ star formation will be suppressed at most radii in the disk.   Taking an NFW profile \citep{NFW} for the halo, 
\begin{eqnarray}
Q(s) = \frac{\alpha}{3.36} \sqrt{\frac{\pi}{G \delta_{\rm c}\rho_{\rm crit}}}
\frac{c_{\rm disp}\Omega_{\rm m}}{r_{\rm s}\Omega_{\rm b}} \times \nonumber \\ 
\frac{s}{s_i s^{1/2}I(s_i)} \frac{ds}{ds_i} 
\sqrt{M(s)+\frac{s^2}{(1+s)^2}}.
\label{eq:Q}
\end{eqnarray}
where $r = s r_s$ and the NFW scale radius $r_s = r_v/c$, where $r_v$ is the virial radius and $c$ the concentration index\footnote{In eq.~\ref{eq:Q} we correct a typographical error in eq.~2 in \citet{darkgal}, namely $df/d\lambda$ should be $ds/ds_i$.} that satisfies $sM(s)=a^2\lambda^2 s_{\rm i} M(s_{\rm i})$, with $a=2.5$.
The mass enclosed is $M(<s)=M_v M(s)/M(c)$ where $M(s) \equiv\ln(1+s)-s/(1+s)$, and the virial mass within $r_v$ is $M_v$. $\delta_{\rm c}\sim 200$ is the characteristic overdensity that parametrises the central density in terms of the critical density $\rho_{\rm crit}$.
The initial surface density is obtained from 
to $|s^2-1|^{3/2}I(s) = \sqrt{s^2-1}-\cos^{-1}(1/s)$ for $s>1$,  and 
$|s^2-1|^{3/2}I(s) =-\sqrt{1-s^2}+\ln\left[s(1-\sqrt{1-s^2})^{-1}\right]$ for $s<1$.  $\Omega_{\rm b,m}$ are the baryon and matter densities respectively.  The velocity dispersion in the disk is taken to be 10 km$\,$s$^{-1}$, which is the value found for galaxies with low star formation rates \citep{KennicuttEvans12,Ianjamasimanana12}, and we fix the concentration index for the haloes to be 20, characteristic of low-mass systems \citep{Wechsler02, White97}. In computing the gravitational force, this formula ignores the fact that the baryons move in as the disk becomes self-supported.  We modify the equation to allow the baryons which settle to the disk and move in to contribute to the gravity vector, ignoring a small correction of this sub-dominant contributor from its distribution not being spherical.  This means solving $(2.5\lambda)^2 s_i M(s_i) = s[(1-\beta)M(s)+\beta M_{\rm proj}(s_i)]$, where $\beta\equiv \Omega_{\rm b}/\Omega_{\rm m}$, and $M_{\rm proj}(s_i)$ is the total mass projected onto the disk, within scaled radius $s_i$.  The results for the minimum spin parameter vs mass are virtually identical.

We combine this with the lognormal distribution of halo spin parameter $\lambda$, with mean 0.04 and dispersion 0.6  as suggested by the most recent N-body simulations \citep{Zurich,Bett}. The Millennium simulation finds nearly 1M halos with very high values of the spin parameter (> 0.2). The authors claim that this is partly due to the very large volume of their simulation, thus they sample the tails of the distribution more efficiently and also to the different choice of halo finder (see their Fig. 1). When they modify their halo finding algorithm they do not find such a tail of high spin halos (see their Fig. 8 and 9) but they still find spin values up to lambda~0.3. This is in better agreement with the Zurich simulations and also with recent analytic considerations that re-analyse the millennium simulation \citep{Benson}. Our derived spin values from observation are all below 0.3 except one point. The N-body simulations show that it is possible to form dark matter halos with such high spin ($\sim 0.3$) in non-dissipative collapse. The spin parameter can be understood on the basis of tidal torque theory of density peaks \citep{HP88}.   The idea is then very simple: larger values of $\lambda$ give rotationally-supported gas disks with larger radii and hence with lower gas surface density, and larger Toomre $Q$ values, and hence are more likely to be stable.  Given that the surface density is not constant with radius, the theory predicts that some disks will be stable in some regions but not others, but \cite{darkgal} showed that disk galaxies with halo masses below about $10^9\,M_\odot$ should be stable throughout, in an Einstein-de Sitter Universe.  Since then, the standard cosmological model has changed, and in this paper we revise the computations for a $\Lambda$CDM  cosmology with a cosmological constant $\Omega_\Lambda = 0.68$, $\Omega_{\rm m} = 0.32$, and $\Omega_{\rm b}=0.045$ ~\citep{Planck18}.

\section{Data and Method}
\label{Method}

Recent HI surveys have found a number of almost-dark-galaxies. These surveys have also measured in great detail the physical properties of these galaxies. For these HI-rich galaxies, the baryonic mass is dominated by the HI component (the typical ratio $M_{*}/M_{\rm HI}$ is $0.03 - 0.1$~\citep{Leisman,Butler} while for the Milky Way the ratio is $10$, so it can be as extreme as $3$ orders of magnitude when compared to a normal galaxy like our own. For the very few stars that are found, the inferred star formation rates can be as low as $10^{-3}$M$_{\odot}$ yr$^{-1}$, again $3-4$ orders of magnitude below the one in the Milky Way. Clearly these are galaxies where star formation is somewhat incidental. For our analysis, we will use publicly-available data from the following surveys, which have measured physical parameters to allow comparison with the model of \citet{darkgal} :

\begin{enumerate}
\item \citet{Leisman} report the discovery of 115 very low optical surface brightness, highly extended, HI-rich galaxies from the ALFALFA survey. They report very low star formation efficiencies. We will analyse the restrictive sample of 30 HI-bearing ultra-diffuse sources (R), with half light radii $r > 1.5 {\rm kpc}$, surface brightness in the $g$ band $\mu_{g,0} > 24$ mag arcsec$^{-2}$ and absolute magnitude   $M_g > -16.8$ mag. These are galaxies with baryonic components dominated by HI. We also analyse the B(road) sample with $r_{\rm eff} > 1.5$ kpc, $\mu(r,r_{\rm eff}) > 24$ mag arcsec$^{-2}$, and $M_r  > -17.6$. This B sample has higher optical  surface brightness  than the R one; galaxies in this sample are also more concentrated and exhibit lower halo spins than the R sample (see Fig.~\ref{fig:radiifit}). They have higher star formation rates and are identified in GALEX, which indicates (mildly more) star formation activity than in the R sample. In any case, as they have significant HI masses, we chose to also include them in our analysis as they are also near dark galaxies.

\item \citet{Butler} report measurements of baryonic mass and specific angular momentum for 14 HI-rich rotating dwarf irregular galaxies in the LITTLE THINGS sample. These are targeted galaxies and not found in a blind survey as above, and, apart from the stability threshold, the distributions we calculate here do not directly apply.
\end{enumerate}

The above samples provide us with $129$ near-dark galaxies with measured values of spin that we can contrast with the predictions in~\citet{darkgal}. The LITTLE THINGS sample has resolved imaging and thus their spin calculations are fairly robust. On the other hand, for the ALFALFA sample, only 3 galaxies have resolved observations. The rest of galaxies are point sources and thus their derived spin values are much more uncertain. This presents some difficulties in comparison, so to assess the robustness of the derived spins, we compute them in two different ways: first using the method in \citet{Leisman} $\lambda = 21.8 R_d / V_{\rm rot}^{3/2}$, where $R_d$ is the scale of the disk in kpc and $V_{\rm rot}$ the flat curve rotation value in km/s. We also estimate the dark halo spin $\lambda$ using the HI mass as $\lambda = 2^{1/2} V_{\rm rot}^2 R_d / (G M_{\rm dark})$, where $M_{\rm dark} = 5.88 M_{\rm HI}$ and 5.88 is the \citet{Planck18} dark matter to baryon universal fraction. When comparing with the predictions in \citet{darkgal} we keep only those galaxies for which both methods give spin parameters that agree to within $0.05$. This results in keeping 30 galaxies (14 from LITTLETHINGS and 16 more from ALFALFA). The low reliability of derived values for $\lambda$ from unresolved sources in the ALFALFA survey is not unexpected due to the lack of spatial resolution.

\begin{figure}
\centering
    \includegraphics[angle=0,clip=,width=\columnwidth]{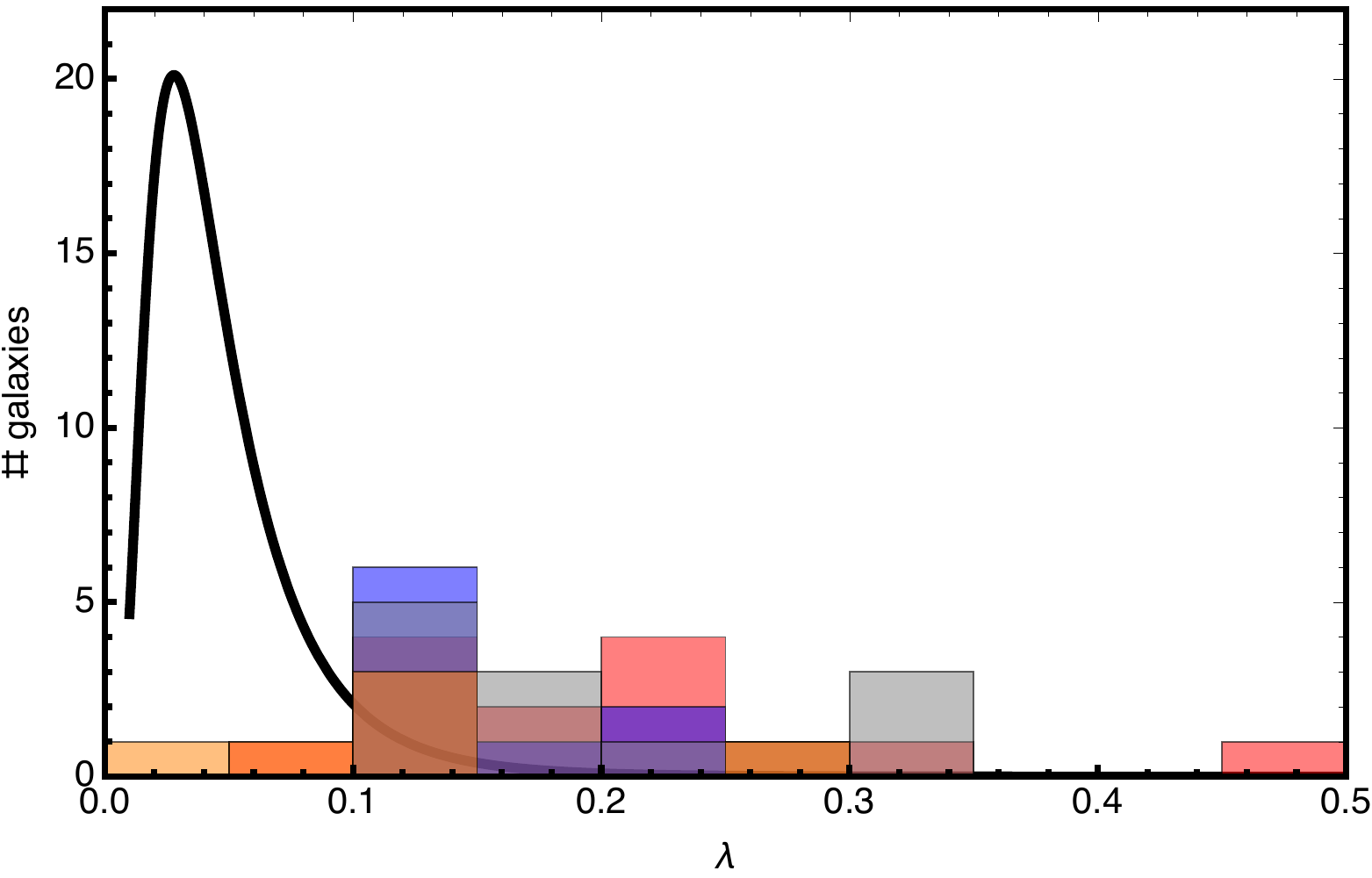}
\caption{Halo spin distribution for galaxies in different samples  and from tidal-torque hierarchical collapse theory (black solid line; this distribution is normalised to unit probability). The red and blue histograms are for galaxies in the ALFALFA R and B samples respectively. The grey histogram corresponds to the LITTLE THINGS sample. The Mancera et al. sample is plotted in orange. The observed galaxies are clearly sampling the tail of the theoretical distribution.}
 \label{fig:spindist}
\end{figure}

\begin{figure}
\centering
    \includegraphics[angle=0,clip=,width=.95\columnwidth]{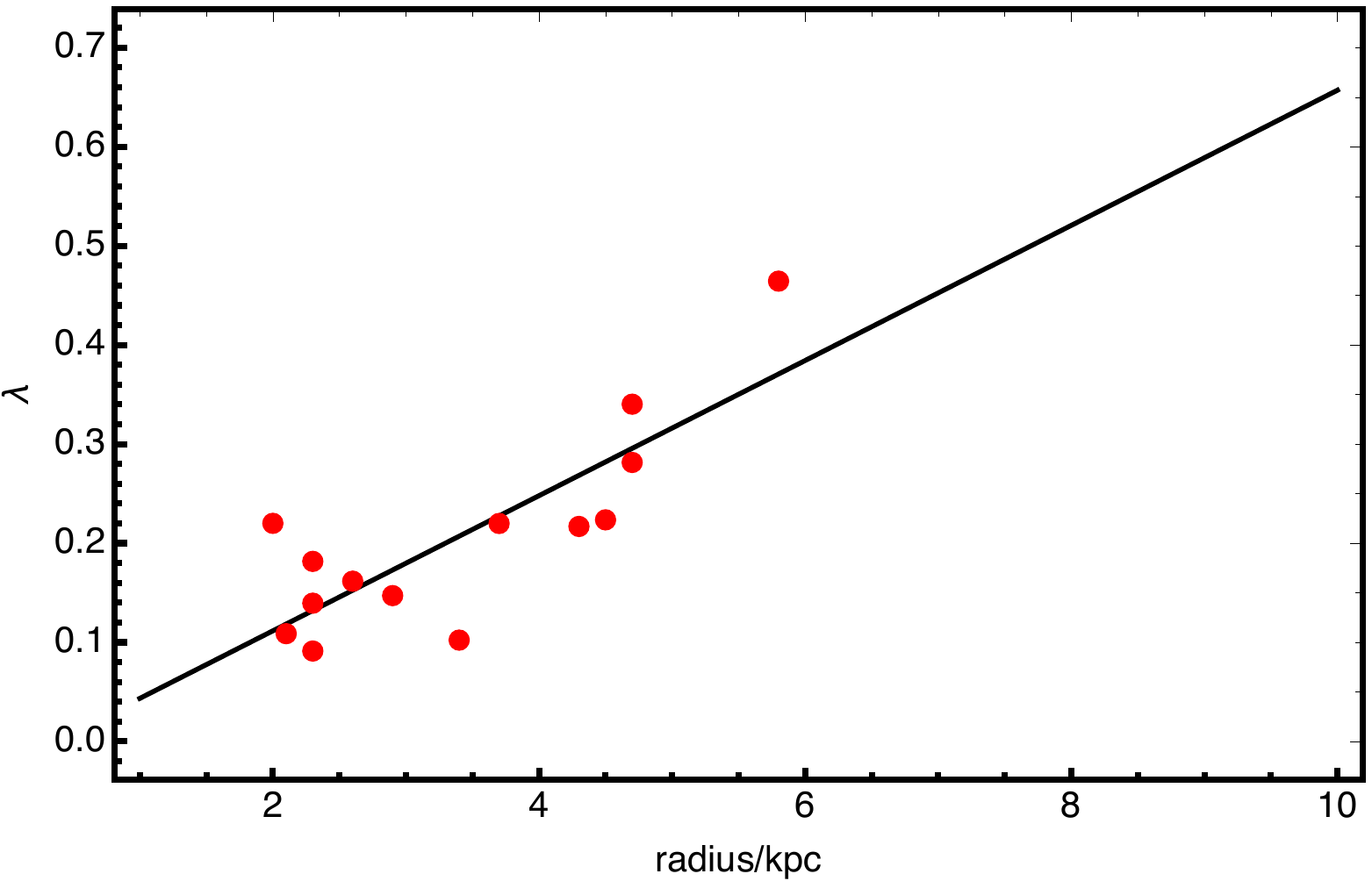}
     \includegraphics[angle=0,clip=,width=.95\columnwidth]{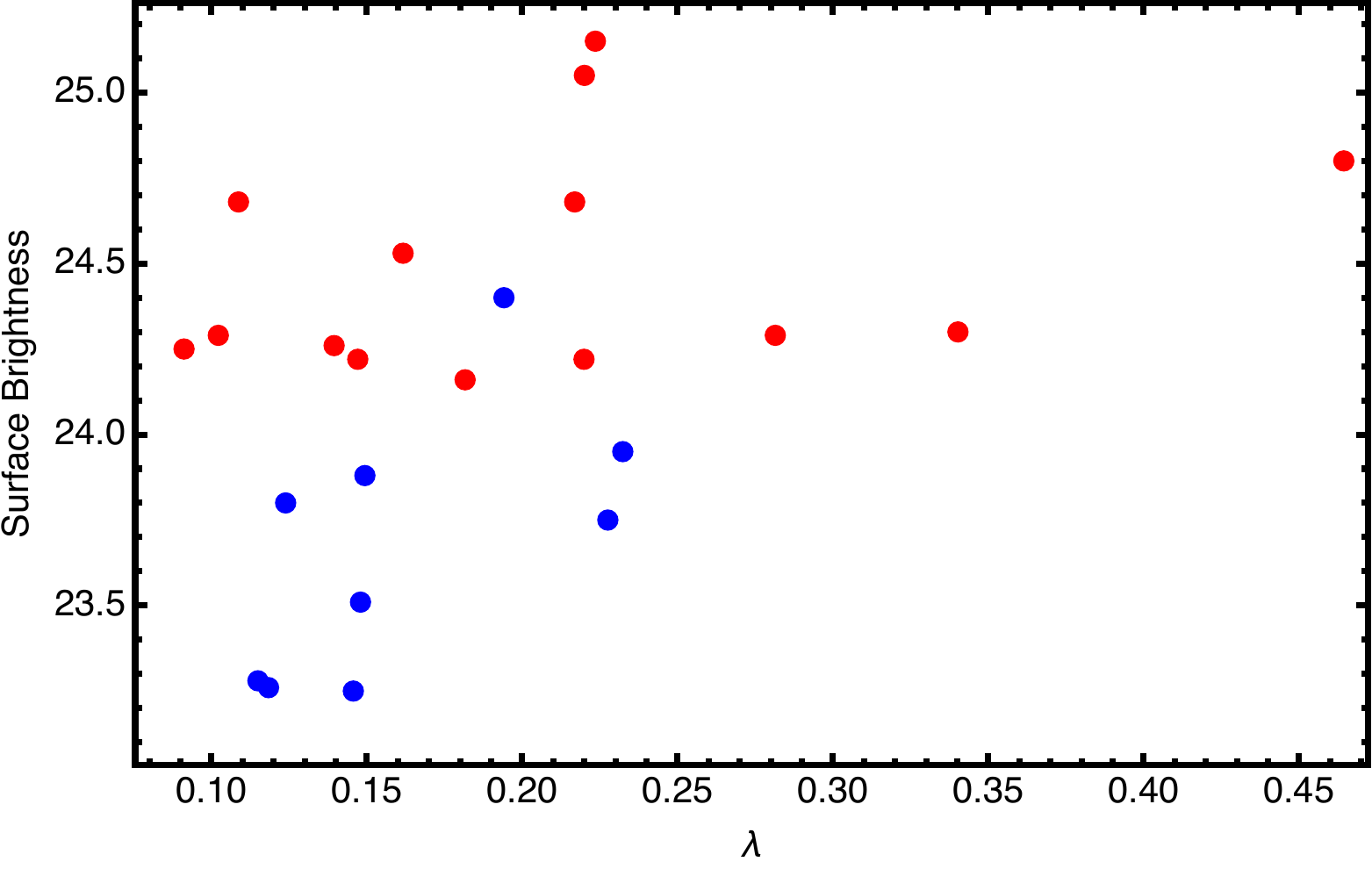}
\caption{Top panel: Halo spin distribution for galaxies in the  \citet{Leisman} sample R (red dots), plotted against optical half-light radius. It shows a strong correlation ($R^2 =0.7$) between $\lambda$ and radius. Bottom panel: Surface brightness vs. $\lambda$ for the R(estricted) and B(road) samples in \citet{Leisman}.  The B(road) sample (blue points) has larger surface density and lower halo spin values (on average) than the R(estricted) sample.}
 \label{fig:radiifit}
\end{figure}

\begin{figure}
\centering
    \includegraphics[angle=0,clip=,width=1.1\columnwidth]{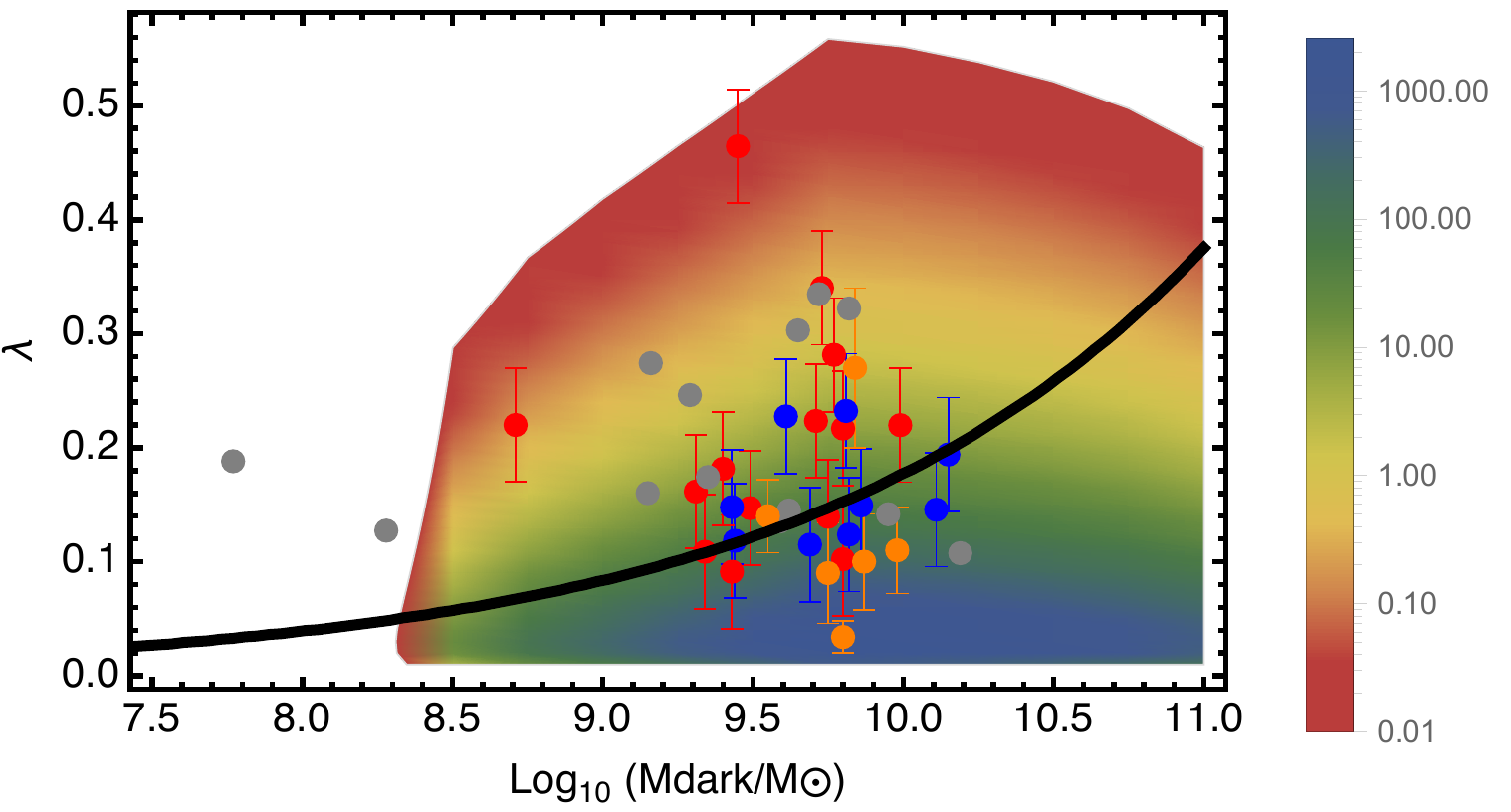}
\caption{Halo Spin as a function of dark halo mass for galaxies in the samples of  LITTLETHINGS (grey points),  R (red) and B (blue) samples from ALFALFA, and from Mancera et al. (orange). The black thick solid line is the tehoretical prediction for the fraction of dark galaxies; this line has no free parameters once the cosmology is chosen. Upward of the theory line all galaxies should be dark; in our model this means that the whole disk is dark. 
The colour density contours show the expected number of galaxies in the ALFALFA sample per 0.1 decade in dark halo mass and $0.01$ in spin value.  Galaxies with halo mass less than about $10^{8.4}$ M$_{\odot}$ are too faint in HI to be in the survey, due to the lower distance limit of 25 Mpc. The majority of (luminous) galaxies in the sample are expected in the blue region. Dark galaxies are expected to be found in the green region, where the abundance is a factor $\sim 100$ lower than for star forming galaxies. The regions in red and yellow are where we expect less than one galaxy in the ALFALFA survey.  Note that the coloured region is only relevant to ALFALFA, as the grey LITTLETHINGS sample is selected differently. We remark on the good agreement of our predictions and observations. Non-dissipative N-body simulations (see text) predict the formation of dark halos with spin up to values of 0.3, in agreement with observations, except for one single point for which we determine $\lambda=0.46$ which could be an outlier, given the difficulty in accurate measurements of the halo spin parameter.}
 \label{fig4}
\end{figure}

\begin{figure}
\centering
    \includegraphics[angle=0,clip=,width=.95\columnwidth]{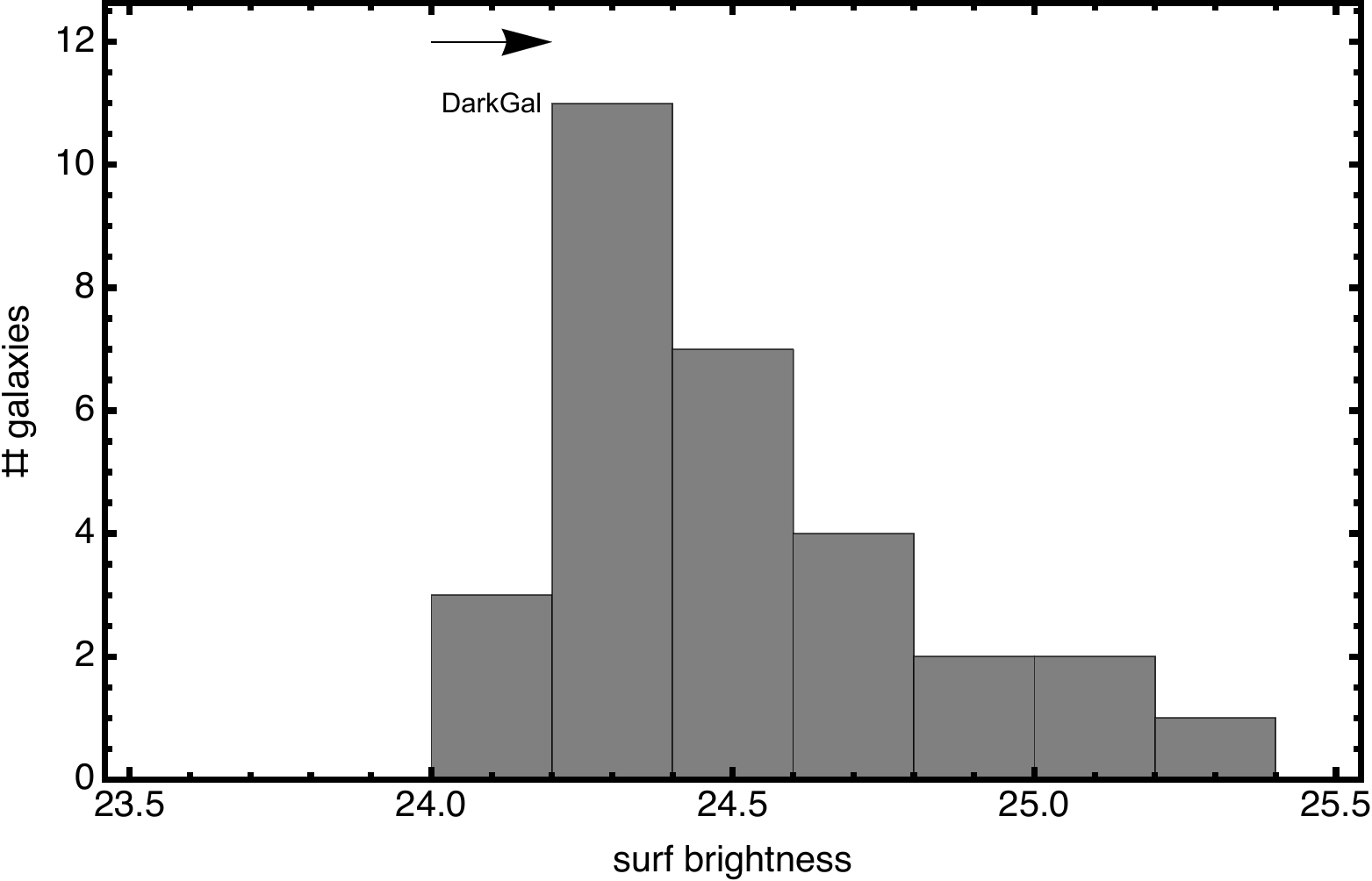}
\caption{Central surface brightness for galaxies in the  \citet{Leisman} sample. The prediction from \citet{darkgal} is the region rightward of the arrowhead. }
 \label{fig5}
\end{figure}

\begin{figure}
\centering
    \includegraphics[angle=0,clip=,width=.95\columnwidth]{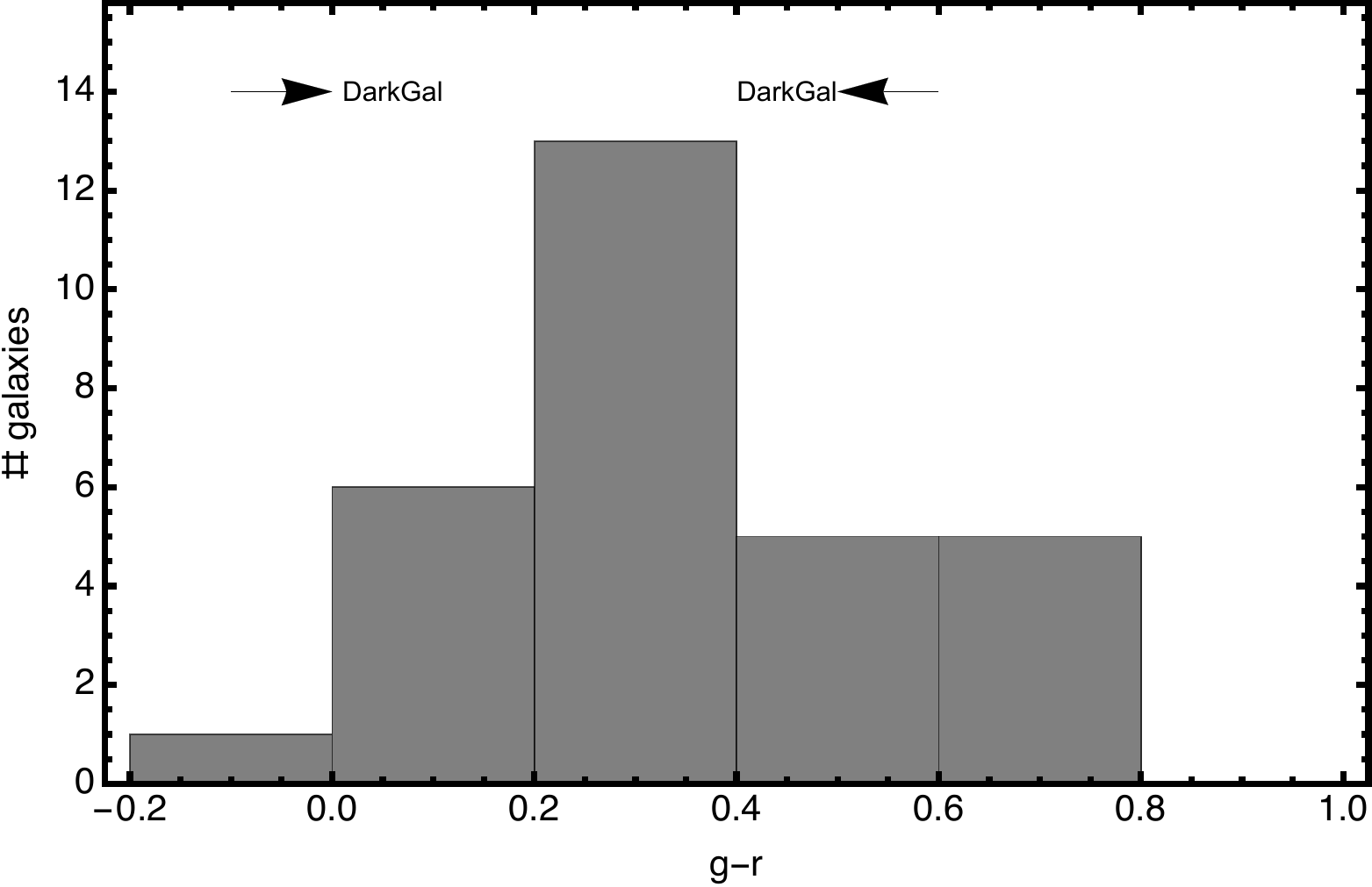}
\caption{$g-r$ colours for galaxies in the  \citet{Leisman} sample. The prediction from \citet{darkgal} is the region between the arroheads. }
 \label{fig6}
\end{figure}

\section{Results}

The distribution of halo spins for the different samples is shown in Fig.~\ref{fig:spindist} for the three different subsamples:  R (red), B (blue) and LITTLETHINGS (gray). We have also plotted the theoretical distribution from tidal torque theory as a black line for mean 0.06 and sigma 0.6 for a log-normal distribution. The theoretical distribution is normalised to the peak of the observed HI galaxies. All HI samples have significant larger halo spins than the theoretical prediction, thus supporting the premise that dark galaxies represent the tail of the spin parameter distribution.

For the R sample we also display in Fig.~\ref{fig:radiifit} (top panel)  the halo spin as a function of radius. There is a correlation between spin and radius, see Fig.~\ref{fig:radiifit}, consistent with the idea that these disks are extended because of the high value of the halo spin. The bottom panel shows the surface brightness versus halo spin for both the R (red dots) and B sample (blue dots). The B sample has higher concentration and lower spins than the R sample. The B sample galaxies are visible in GALEX and are less dark than the R sample. They have more star formation and we expect the B sample to have potentially galaxies below the threshold for star formation.

Fig.~\ref{fig4} shows the distribution of spin as a function of halo mass for the  R(red dots), B(blue dots) and the LITTLETHINGS samples (grey dots). For the R and B samples we have kept only those galaxies for which the $\lambda$ values agrees to $0.05$ in the two methods, as explained in \S~3. The solid black line is the prediction from \citet{darkgal}, above which no stars are expected to form at any radius in the disk, i.e. the disk at all radii has $Q > 1$. The change to $\Lambda$CDM  cosmology reduces the minimum mass for galaxies to be dark.  We remark that this line has no free parameters once the cosmology is chosen

Six of the galaxies in the \cite{Leisman} sample have been observed by \cite{Mancera2019}, who find that they are outliers in the Tully-Fisher relation, and consistent with no baryons having been ejected.    We also plot these galaxies in Fig.~\ref{fig4}.
 
There is good agreement with the prediction from \citet{darkgal}; firstly, given the uncertainties in the spin parameters, the points are consistent with being above the line.  The colour density contours show the expected number of galaxies, using the appropriate survey geometry, and deriving the mass-dependent volume limits from \citet{Haynes} for the ALFALFA sample.  The numbers shown are per 0.1 decade in dark halo mass and per $0.01$ in spin value, using the baryonic mass function from~\cite{Panter07}. We see that the majority of dark ALFALFA galaxies are in the high-density green region, and exist in far smaller numbers than the regular galaxies that occupy the blue regions below the line. The abundance density in the green regions is a factor $\sim 100$ lower than for star forming galaxies. The region in red is where we expect less than one galaxy in the ALFALFA survey. Note that the colouring is only relevant to ALFALFA, as the grey LITTLETHINGS sample is selected differently.  Large survey volumes are needed to find the dark galaxies with very high spin parameter and dark halo mass above $5 \times 10^{10}$ M$_{\odot}$ as their expected abundance is below one for the ALFALFA survey volume. Note that we expect most galaxies in the survey to be luminous and with low spins (blue region). To find more dark galaxies it is better to cover larger volumes (the red region is for expectation of 0.1 dark galaxies in the ALFALFA sample) and go to lower masses, which ALFALFA is not sensitive to~\citep{Leisman}. We remark on the good agreement of our predictions and observations. B sample galaxies tend to be slightly closer to  the theoretical line for stability from our model. This is expected as they are more concentrated than the R sample. Their surface brightness is slightly below our predicted cut-off for galaxies to be dark. But overall those galaxies in the B sample with robust derived values for $\lambda$ follow our prediction within the uncertainty.

Fig.~\ref{fig5} and~\ref{fig6} show the photometric properties of the \citet{Leisman} sample for the  central surface brightness in $r$ and $g-r$ colour. Shown is also the range of our predictions for those quantities from \citet{darkgal}. We have converted our prediction in Fig.~5 and 6 from \citet{darkgal} from Johnson to SDSS filters using the transformations in \citet{Rodgers} and applied the K-corrections from \citet{Sodre} to bring them to the z of the \citet{Leisman} sample ($z \sim 0$). Fig.~\ref{fig5} shows that the predicted surface brightness matches well with the observed one. A similar story can be found for the colours of these galaxies; Fig.~\ref{fig6} shows that the predicted colour for galaxies corresponds to blue galaxies, consistent with sporadic activity of small bursts of star formation.

\subsection{Number density of high-spin haloes}

Is the observed number of HI galaxies consistent with being the tail of the theoretical spin distribution? The HI surveys cover a volume of $8 \times 10^5$ Mpc$^{3}$ assuming a Planck18 cosmology (this is $1/8$th of the sky for a proper distance between $25$ and $120$Mpc). We use the baryonic mass function from~\cite{Panter07} with parameters for a Schechter function $\phi_* = 2.2 \times 10^{-3}$ Mpc$^{-3}$, $M_* = 1.005 \times 10^{11}$ M$_{\odot}$ and slope $\alpha = - 1.22$ and the Planck18 $\Omega_m/\Omega_b = 5.88$ ratio, the expected number of total galaxies in the above volume for the dark halo mass range $10^{9} - 10^{10.2}$ M$_{\odot}$ is $n_{\rm gal} = 8 \times 10^3$ galaxies.

From the theoretical spin distribution, we expect 26 galaxies above $\lambda=0.3$, and the surveys see 22.  However, two caveats are in order: the probabilities are so low that we are extrapolating the spin distribution to higher spins than where it is reliably determined from simulations; and at such high spins, the NFW halo may well not be well-approximated as spherical.

We conclude that the observed number of dark galaxies in HI surveys is consistent with them being the highest spins from the theoretical distribution from hierarchical galaxy formation.  It also appears that the HI surveys have found nearly all the dark galaxies in the relevant mass range, and we do not expect to find large numbers more dark galaxies above a dark halo mass of $10^{10}$M$_{\odot}$.  There should be lower mass dark galaxies, which a lower HI flux limit should reveal\footnote{The ALFALFA~\citep{Leisman} selection criterion for the R and B samples implies HI masses above $10^8$ M$_{\odot}$ ($6 \times 10^8$ M$_{\odot}$ in dark matter) due to the requirement of isolation. Alternatively, the mass limit can be deduced directly from the data presented in \cite{Haynes} Fig. 3.}. Here our model predicts that all disks should be dark (see also~\citet{VerdeOh}). For dark halo masses above $3 \times 10^{10}$ M$_{\odot}$ we predict that there should be dark galaxies only above $\lambda > 0.25$, which are extreme values for the spin halo. There should be $10^{-6}$ Mpc$^{-3}$ of these, so a larger volume than the current ALFALFA volume is needed to have a good chance of  finding one of these.


\section{Summary}

We have compared the observed properties of the almost-dark galaxies, recently found in local HI surveys, to the predictions of the model proposed in~\citet{darkgal}. In this model we proposed that disk galaxies with high spin would be dark as they would be Toomre stable. This led us to predict a population of dark galaxies. Here we have updated our predictions to the $\Lambda$CDM cosmology and compared to the spins, masses and number densities of the observed dark galaxies. While all the HI-rich galaxies do contain some stars, these are highly subdominant to the baryonic mass budget of the galaxies ($< 10$\%). Furthermore, their star formation rates are also $3-4$ orders of magnitude below that of the Milky Way (which is itself not a very vigorous star forming disk).  Therefore, while acknowledging that the theory is not a perfect description, omitting for example star formation activity that might arise from a small central bulge, we compare the observed almost-dark galaxies with disks that the theory indicates should be completely stable.

We have shown that the mass and spin distribution of the  observed dark galaxies is in agreement with our prediction.  Although measuring the spin parameter for unresolved galaxies is challenging, they appear to inhabit the region that is Toomre stable, and moreover preferentially the part of this region that has the highest expected number density. The colours and surface brightness of dark galaxies are also consistent with predictions. We have also shown that the number density of the HI dark galaxies is consistent with them being the tail of the spin parameter distribution predicted from hierarchical tidal torque theory. The predicted and observed abundance agree, so we predict that significant numbers of new dark galaxies would only be found at lower masses (in fainter HI samples) or in small number at higher masses (up to $\sim 3 \times 10^{10}$ M$_\odot$, and extremely high spin; see green region in Fig.~\ref{fig4}). This would be consistent with what is inferred from weak lensing surveys regarding the abundance of low mass galaxies~\citep{JimenezWL} in the cosmological volume.

The HI near-dark galaxies, as well as the three dark galaxies in \citep{Leisman2}, are all chosen to be located in isolated environments with over-densities nearby. This is in line with the predictions of tidal-torque theory that high-spin halos will be in the low-density regions. \citet{Berta} find an anti-correlation between halo mass and spin in SDSS galaxies which is not seen in the dark galaxies, and find lower spin values than the HI samples, as expected. Our findings imply that there is a spin bias in galaxy formation similar to the peak bias in \cite{Kaiser} but here arising from the spin parameter of the halo influencing the stability of the disk to star formation.

\section*{Acknowledgments}

RJ acknowledges support by Spanish MINECO grant PGC2018-098866-B- I00.  We thank Alex Geringer-Sameth and the anonymous referee for useful comments.

\section*{Data Availability}
The data underlying this article are available in the article and references therein.

\end{document}